\documentclass[aps,prd,twocolumn,superscriptaddress,nofootinbib,10pt]{revtex4-1}
\usepackage{amsmath,amssymb,bm,latexsym,acronym,float,soul,nicefrac,verbatim,cancel,accents}
\usepackage{multirow,dcolumn,longtable,footnote,appendix,graphicx,color,subfigure}
\usepackage[colorlinks,linkcolor=blue,citecolor=blue,urlcolor=blue ]{hyperref}
\usepackage[normalem]{ulem}

\newcommand{\TRC}{MOE Key Laboratory of TianQin Mission, TianQin Research Center for Gravitational Physics $\&$ School of Physics and Astronomy, Frontiers Science Center for TianQin, Gravitational Wave Research Center of CNSA, Sun Yat-sen University (Zhuhai Campus), Zhuhai 519082, China}

\allowdisplaybreaks
\newcommand{\be}{\begin{equation}}
\newcommand{\ee}{\end{equation}}
\newcommand{\bea}{\begin{eqnarray}}
\newcommand{\eea}{\end{eqnarray}}
\newcommand{\nn}{\nonumber}

\newcommand{\cF}{{\cal F}}

\newcommand{\cK}{{\cal K}}

\newcommand{\cP}{{\cal P}}

\newcommand{\pd}{\partial}
\newcommand{\td}{\tilde}
\newcommand{\wtd}{\widetilde}

\newcommand{\sbs}{\rule{2.5pt}{2.5pt}}

\begin{document}

\title{\bf Separating the linearized Einstein equations in Kerr}

\author{Jianwei Mei}
\email{Email: meijw@sysu.edu.cn}
\affiliation{\TRC}

\begin{abstract}
A direct separation of the linearized Einstein equations in Kerr is presented.
The trace of the perturbed metric is found to obey the spin-0 perturbation equation and so is irrelevant for the spin-2 perturbations.
By combining the traceless condition, the Killing-Yano symmetry, a parity requirement, and the de Donder gauge condition, it has been found possible to reduce the linearized Einstein equations in Kerr to a manageable level, thereby enabling the direct separation of the equations and the derivation of a unique set of decoupled mode functions for the spin-2 perturbation of the Kerr black hole.
\end{abstract}

\maketitle



\acrodef{GR}{general relativity}
\acrodef{GW}{gravitational wave}
\acrodef{EMRI}{Extreme Mass Ratio Inspiral}
\acrodef{MBHB}{massive black hole binary}
\acrodef{BHPT}{Black hole perturbation theory}
\acrodef{LEE}{linearized Einstein equation}
\acrodef{DCS}{Differentiate-Cancel-Simplify}
\acrodef{KY}{Killing-Yano}

\section{Introduction}\label{sec:intro}

A long-standing problem in black hole perturbation theory is to separate the \acp{LEE} in the background of a Kerr black hole,
\bea\delta G_{\mu\nu}=8\pi G_N \td{T}_{\mu\nu}\,,\label{eq.LEE0}\eea
where $\delta$ generates the perturbation, $g_{\mu\nu}\; \longrightarrow\; g_{\mu\nu} + \delta g_{\mu\nu}$, $\delta G_{\mu\nu}$ is the linearized Einstein tensor, and $\td{T}_{\mu\nu}$ is the corresponding source.

The most logically straightforward procedure is first to reduce the ten coupled partial differential equations in \eqref{eq.LEE0} to a single master equation in one unknown function (the master function), and then to separate this master equation into a set of ordinary differential equations.
To cope with the gauge freedom inherent in \eqref{eq.LEE0}, one may simultaneously impose a gauge-fixing condition such as the de Donder gauge,
\bea\nabla^\alpha h_{\alpha\beta}-\frac12\pd_\beta h=0\,,\label{eq.gauge.DD}\eea
where $h_{\mu\nu}\equiv\delta g_{\mu\nu}\,$ and $h\equiv g^{\mu\nu}h_{\mu\nu}$.
To reduce the number of equations, one needs to solve for nine of the ten perturbed metric components, $h_{\mu\nu}\,$, from \eqref{eq.LEE0} and \eqref{eq.gauge.DD} sequentially.
In order not to lose any generality, the only feasible way appears to be via an iteration of the \ac{DCS} procedure.\footnote{Suppose a single-variable function $f(r)$ is differentiated to orders $n$ and $k$ (with $k>n$) in two equations,
\begin{subequations}
\begin{align}
&\cdots+\pd^{n}_rf=0\,,\label{eq.DES1a}\\
&\cdots+\pd^{k}_rf=0\,,\label{eq.DES1b}
\end{align}
\end{subequations}
then one can differentiate \eqref{eq.DES1a} $k-n$ times with respect to $r$, using the resulting expression to cancel the $\pd^{k}_rf$ term in \eqref{eq.DES1b}, and then simplify.
By iterating this \ac{DCS} procedure, one can reduce the derivative order of $f$ to zero under suitable conditions, enabling an algebraic solution for $f$ without any loss of generality.
The idea is the same for multi-variable functions in partial differential equations.}
However, due to the complexity of the equations, the process quickly becomes computationally infeasible after the elimination of only a few perturbed metric components.
As a result, the above process has never been successfully carried out for the \acp{LEE} in Kerr.

This problem is not merely a technical nuisance --- it has become a pressing obstacle for taking advantage of the ever growing \ac{GW} data.
Obtaining the linear order perturbation is the first step to the study of nonlinear effects in black hole perturbation theory.
This has become particularly important as resolvable nonlinear effects are not only promising targets for future space-based \ac{GW} detectors \cite{Shi:2024ttu,Luo:2025ewp,LISA:2022kgy} but also possibly detectable with current ground-based detectors \cite{Yang:2025ror,Wang:2026rev}.
The lack of a simple method to obtain the linear order metric perturbation in Kerr has become an apparent obstacle for the precise modeling of some types of \ac{GW} signals, such as for the modelling of self-force effect \cite{Poisson:2011nh,Barack:2018yvs,Pound:2021qin} for the \ac{EMRI} signals \cite{Poisson:2011nh,Barack:2018yvs,Pound:2021qin,Berens:2024czo,Hollands:2024iqp}.

Existing work addressing the problem mainly includes the following.
\begin{itemize}
\item Metric reconstruction.
In the early 1970s, Teukolsky discovered the separability of the equations for the perturbed Weyl scalars $\psi_0$ and $\psi_4$ \cite{Teukolsky:1972my,Teukolsky:1973ha}.
Given $\psi_0$ or $\psi_4$, it is possible to recover the perturbed metric through metric reconstruction \cite{Chrzanowski:1975wv,Kegeles:1979an,Wald:1978vm,Stewart:1978tm}.
However, the procedure is not so straightforward:
to obtain the perturbed metric,
one has to start by solving for one of the perturbed Weyl scalars ($\psi_0$ or $\psi_4$),
using it to solve for a Hertz potential,
and then use the Hertz potential to generate the perturbed metric.
The resultant metric is ambiguous up to a piece from perturbing the background charge, and, due to the reliance on the radiation gauge, is also marred by gauge singularities in the presence of a source.
For detailed discussions on the topic, see, e.g. \cite{Berens:2024czo,Hollands:2024iqp} and references therein.
\item Coupled spectral analysis.
Instead of trying to separate the equations, one can try to work with known but coupled function bases, such as by using the spherical harmonics to decompose the angular dependence of the perturbed metric.
Such an idea has been successfully implemented for rotating black holes in \ac{GR} and beyond  \cite{Chung:2023zdq,Chung:2023wkd,Chung:2024vaf,Chung:2024ira}.
However, since the function bases are coupled to each other, one has to truncate the infinite tower of coupled modes at a chosen order and the amplitudes and spectra of the remaining modes have to be solved numerically.
\item Special limits.
The problem becomes easier to tackle when the spin of the background black hole takes limiting values, such as working in the near-horizon extremal Kerr background \cite{Chen:2017ofv} or in the small rotation limit \cite{Franchini:2023xhd}.
\end{itemize}
In view of the limitations of existing methods, it is of fundamental importance to separate the \acp{LEE} in Kerr directly, with which one can obtain the perturbed metric directly once the master function is known.

In \cite{Mei:2023pho}, an effort had been made to separate the \acp{LEE} in Kerr by employing their symmetries more extensively.
A new symmetry operator had been constructed from the \ac{KY} tensor, enabling the direct construction of two sets of decoupled mode bases (called solutions) for the spin-2 perturbation of the Kerr black hole.
However, it still had not been possible in \cite{Mei:2023pho} to reduce the ten coupled partial differential equations in \eqref{eq.LEE0} to a single master equation of only one unknown function.

In this work, the task is finally achieved by incorporating two additional features of the \acp{LEE}.
\begin{itemize}
\item The trace of the perturbed metric is found to obey the spin-0 perturbation equation.
So, the spin-2 part of the metric perturbation must be traceless.
\item The \acp{LEE} have a parity symmetry.
So, one can focus on the metric perturbations with a fixed parity.
\end{itemize}
Together with the \ac{KY} symmetry \cite{Mei:2023pho} and the de Donder gauge condition, these new constraints can be used to help reduce the \acp{LEE} in Kerr to a manageable level, thus enabling the direct separation of the equations.
In the process, the perturbed metric will be explicitly expressed in terms of some master functions for the first time.

The paper is organized as following.
In section \ref{sec:eqns}, all the relevant equations and constraints are introduced.
In section \ref{sec:methd}, the key steps for reducing and separating the equations are explained in detail.
In section \ref{sec:rst}, the key results on the perturbed metric components are presented.
In section \ref{sec:sum}, the paper concludes with a brief summary and discussion.

\section{Equations and Constraints}\label{sec:eqns}

As a proof-of-principle demonstration, only the vacuum case will be considered in this work.
In this case, the \acp{LEE} can be written as
\bea \wtd\Box [h_{\mu\nu}]&\equiv&\delta R_{\mu\nu}\nn\\
&=&\frac12\nabla^\rho(\nabla_\mu h_{\nu\rho}+\nabla_\nu h_{\mu\rho}-\nabla_\rho h_{\mu\nu})-\frac12\nabla_\mu\nabla_\nu h\nn\\
&=&0\,,\label{eq.huv}\eea
where $R_{\mu\nu}$ is the Ricci tensor, and $\wtd\Box$ is the wave operator.
The covariant derivative $\nabla_\mu$ is defined with respect to the background Kerr metric $g_{\mu\nu}$, which, in the Boyer-Lindquist coordinates $x^\mu\in\{t,r,x\equiv\cos\theta,\phi\}$, can be written as
\bea ds^2&=&-dt^2+H\Big(\frac{dr^2}{X}+\frac{dx^2}Y\Big)+(r^2+a^2)Yd\phi^2\nn\\
&&+\frac{2Mr(dt-aYd\phi)^2}{H}\,,\label{metric.kerr}\eea
where $X=r^2+a^2-2Mr\,$, $Y=1-x^2$, $H=r^2+a^2x^2\,$.

In \cite{Mei:2023pho}, a symmetry operator has been constructed from the \ac{KY} tensor $k_{\mu\nu}$,
\bea\cK_4[h_{\mu\nu}]&\equiv&(k^{\alpha\rho} k^\beta_{~\nu} -k^{\beta\rho} k^\alpha_{~\nu})\nabla_\alpha\nabla_\beta h_{\rho\mu}\nn\\
&&+(k^{\alpha\rho} k^\beta_{~\mu} -k^{\beta\rho} k^\alpha_{~\mu})\nabla_\alpha\nabla_\beta h_{\rho\nu}\nn\\
&&-k^{\beta\rho}(k_{\alpha\rho}\nabla_\beta\nabla^\alpha h_{\mu\nu}+\nabla_\beta k_{\alpha\rho} \nabla^\alpha h_{\mu\nu})\nn\\
&&+\nabla^\alpha k^{\beta\rho}(h_{\alpha\mu}\nabla_\nu k_{\beta\rho}+ h_{\alpha\nu}\nabla_\mu k_{\beta\rho})\nn\\
&&-h^{\alpha\beta}(\nabla_\mu k_{\alpha\rho} \nabla_\nu k_\beta^{~\rho}
+\nabla_\nu k_{\alpha\rho}\nabla_\mu k_\beta^{~\rho})\nn\\
&&+2(J_\nu^{~\alpha\beta}-J^{\alpha~\beta}_{~\nu})\nabla_\beta h_{\alpha\mu}\nn\\
&&+2(J_\mu^{~\alpha\beta}-J^{\alpha~\beta}_{~\mu})\nabla_\beta h_{\alpha\nu}\,,\eea
where $J^\alpha_{~\mu\nu}=k_{\rho\mu}\nabla^\alpha k^\rho_{~\nu} -k_{\rho\nu}\nabla^\alpha k^\rho_{~\mu}\,$.
Since $\cK_4$ commutes with $\wtd\Box$ off-shell, it can be used to help classify the solutions to \eqref{eq.huv}.
This can be achieved by looking for simultaneous solutions to \eqref{eq.huv} and the eigen-equation,
\bea\cK_4[h_{\mu\nu}]=\lambda h_{\mu\nu}\,,\label{eq.eigen.guv}\eea
where $\lambda$ is a constant.

Due to the time translation and rotation symmetry of the background Kerr spacetime, the $(t,\phi)$-dependence of $h_{\mu\nu}$ can be easily separated through the mode decomposition,
\bea h_{\mu\nu}(t,r,x,\phi)=\sum_{\lambda wm}A_{\lambda wm} h^{(\lambda wm)}_{\mu\nu}(t,r,x,\phi)\,,\label{mode.decomp.huv}\eea
where $h^{(\lambda wm)}_{\mu\nu}(t,r,x,\phi)= e^{-i(wt-m\phi)}f^{(\lambda wm)}_{\mu\nu}(r,x)$ is the mode basis, and $A_{\lambda wm}$ is the amplitude of each mode.
To align with the prevailing convention in the literature, the signs of $w$ and $m$ have been reversed with respect to those in \cite{Mei:2023pho}.

For decoupled modes, each $h^{(\lambda wm)}_{\mu\nu}(t,r,x,\phi)$ is an independent solution to the equations \eqref{eq.huv}, \eqref{eq.gauge.DD} and \eqref{eq.eigen.guv}.
So, one can focus on a single mode and drop the superscript $(\lambda wm)$ from the mode function $f^{(\lambda wm)}_{\mu\nu}(r,x)$.
In this way, one can get from \eqref{eq.huv}, \eqref{eq.gauge.DD} and \eqref{eq.eigen.guv} a set of 24 equations for the ten mode functions $f_{\mu\nu}(r,x)$, which can be written schematically as
\bea E^{(24)}\Big[f_{\mu\nu}(r,x)\Big]=0\,.\label{eq.combine24}\eea
Here and below, functionals like $E^{(n)}[\cdots]$,  $\cF^{(n)}_{\mu\nu}[\cdots]$, and $\td\cF^{(n)}_{\mu\nu}[\cdots]$ will be used to schematically denote a set of $n$ partial differential equations.
All these equations are too complicated to be presented in a paper.
Fortunately, such details are not need for understanding the basic ideas of the work.

Because the perturbed Weyl scalars $\psi_0$ and $\psi_4$ are known to satisfy Teukolsky's master equation, it will be helpful to treat $\psi_0$ and $\psi_4$ as known functions and use their definition as extra constraints on the perturbed metric components,
\bea \psi_0=\psi_0[h_{\mu\nu}]\,,\quad \psi_4=\psi_4[h_{\mu\nu}]\,,\label{eq.weyle.scalar}\eea
where the details of the operators $\psi_0[h_{\mu\nu}]$ and $\psi_4[h_{\mu\nu}]$ can be found in \cite{Chrzanowski:1975wv}.
In doing this, one can take $\psi_0$ and $\psi_4$ to be the master functions, and all the perturbed metric components will be solved in terms of them.

With the following mode decomposition,
\bea \psi_0&=&\sum_{\lambda wm}A_{\lambda wm} e^{-i(wt-m\phi)}f^{(\lambda wm)}_0(r,x)\,,\nn\\
\psi_4&=&\sum_{\lambda wm}A_{\lambda wm} e^{-i(wt-m\phi)}f^{(\lambda wm)}_4(r,x)\,,\label{mode.decomp.psi04}\eea
one can obtain from \eqref{eq.weyle.scalar} two constraint equations for each mode, which can be written schematically as
\bea \cF_0[f_{\mu\nu}(r,x)]&=&f_0(r,x)\,,\nn\\
\cF_4[f_{\mu\nu}(r,x)]&=&f_4(r,x)\,,\label{eq.weyle.scalar2}\eea
where the superscript $(\lambda wm)$ has also been dropped.

Combining \eqref{eq.weyle.scalar} and \eqref{eq.weyle.scalar2}, one has for each mode 26 equations in total.
However, they are still not enough for one to reduce them to a single equation of only one unknown function.
This situation can be fundamentally changed by incorporating two additional features of the \acp{LEE} in Kerr.

Firstly, by taking the trace of \eqref{eq.huv} and then using \eqref{eq.gauge.DD}, one can find:
\bea g^{\mu\nu}\wtd\Box [h_{\mu\nu}]=-\frac12\nabla_\mu\nabla^\mu h=0\,.\eea
It is obvious that the trace $h$ obeys the equation of a spin-0 perturbation.
So, the spin-2 part of the metric perturbation must be traceless, i.e.,
\bea h=0\,.\label{eq.traceless}\eea
One can check that the two solutions found in \cite{Mei:2023pho} are indeed traceless.
The explicit metric reconstruction in \cite{Berens:2024czo} was also performed under the traceless assumption.

Secondly, by introducing the parity operator,
\bea \cP[f(r,x)]\equiv f(r,-x)\,,\quad \forall\; f(r,x)\,,\label{def.parity}\eea
one can check that the parity conjugate of \eqref{eq.combine24} is still valid,
\bea \cP E^{(24)}\Big[f_{\mu\nu}(r,x)\Big]=0\,,\eea
given that all the mode functions $f_{\mu\nu}(r,x)$ have a fixed parity, i.e.,
\bea f_a(r,-x)=\epsilon f_a(r,x)\,,\quad f_b(r,-x)=-\epsilon f_b(r,x)\,. \label{def.parity.fuv}\eea
Here $\epsilon=\pm1\,$,
$a\in\{{tt}$, ${tr}$, ${t\phi}$, ${rr}$, ${r\phi}$, ${xx}$, ${\phi\phi}\}\,$, and
$b\in\{{tx}$, ${rx}$, ${x\phi}\}\,$.
In the following, $\epsilon$ will be kept as a free parameter.
Solutions with $\epsilon=1$ ($-1$) will be said to have even (odd) parity.

Neither $\psi_0$ nor $\psi_4$ has a fixed parity under the assumption of \eqref{def.parity.fuv}.
As a result, one can get two more constraint equations by acting on \eqref{eq.weyle.scalar2} with the parity operator,
\bea\cP\cF_0[f_{\mu\nu}(r,x)]=f_0(r,-x)\,,\nn\\
\cP\cF_4[f_{\mu\nu}(r,x)]=f_4(r,-x)\,.\label{eq.weyle.scalar3}\eea
On a computer, it will be convenient to introduce
\bea\td{f}_0(r,x)&\equiv&\epsilon f_0(r,-x)\,,\nn\\
\td{f}_4(r,x)&\equiv&\epsilon f_4(r,-x)\,.\label{eq.weyle.scalar4}\eea

\section{Steps to reduce and separate the equations}\label{sec:methd}

In this section, the key steps for reducing and separating the equations will be explained in detail.

\subsection{Warmup}

To see the usefulness of \eqref{eq.traceless} and \eqref{eq.weyle.scalar3}, one can start with \eqref{eq.combine24} and \eqref{eq.weyle.scalar2} only, which are the ones studied in \cite{Mei:2023pho}.

\subsubsection{Reducing the order of $r$-derivatives}

To reduce the number of equations, one can firstly try to lower the order of derivatives with respect to one of the coordinates, say $r$.
On a computer, this can be achieved by iterating the \ac{DCS} procedure without much difficulty.
The results can be written schematically as
\begin{subequations}\label{eq.key.1a}
\begin{align}
&\dot{f}_{\mu\nu}=\cF^{(10)}_{\mu\nu}\Big[f_{\sbs\,\sbs}, f'_{\sbs\,\sbs}, f''_{\sbs\,\sbs}, f'''_{\sbs\,\sbs} \Big]\,,\label{eq.k1a1}\\
&0=E^{(16)}\Big[f_{\sbs\,\sbs},f'_{\sbs\,\sbs},\cdots,f^{(5)}_{\sbs\,\sbs},f_{\sbs}, f'_{\sbs}, f''_{\sbs} \Big]\,,\label{eq.k1a2}
\end{align}
\end{subequations}
where $f_{\mu\nu}, f_{\sbs\,\sbs} \in\{f_{tt}$, $f_{tr}$, $f_{tx}$, $f_{t\phi}$, $f_{rr}$, $f_{rx}$, $f_{r\phi}$, $f_{xx}$, $f_{x\phi}$, $f_{\phi\phi}\}$, $f_{\sbs} \in\{f_0, f_4\}$, $\dot{f}\equiv\pd_r f$, $f'\equiv\pd_x f$, $\cdots$, $f^{(5)}\equiv\pd^5_x f$, and so on.
In this way, the 26 equations in \eqref{eq.combine24} and \eqref{eq.weyle.scalar2} are reduced to 10 equations in \eqref{eq.k1a1} and 16 in \eqref{eq.k1a2}.

\subsubsection{Reducing the order of $x$-derivatives}

The equations in \eqref{eq.k1a2} can be further reduced by lowering the order of derivatives with respect to $x$, again by using the \ac{DCS} procedure.
At this stage, however, each iteration of the \ac{DCS} procedure often takes a too long time to finish.
To accelerate the calculation, one can give numerical values to the physical parameters, $M$, $a$, $w$, $m$ and $\lambda$.
In this way, one can greatly reduce the computation time for each iteration.
In the end, one can reduce \eqref{eq.k1a2} to 12 equations, among which ten are of the form,
\bea f''_{\mu\nu}&=&\td\cF^{(10)}_{\mu\nu}\Big[f_{\sbs\,\sbs},f'_{\sbs\,\sbs}, f_{\sbs}, f'_{\sbs}\Big]\,,\label{eq.key.1b}\eea
which have at most second-order derivatives with respect to $x$, and the remaining two are:
\bea f''_0&=&\frac{2x}Y f'_0+\Big[w^2a^2-\frac{\lambda+2wa(m-2x)}Y\nn\\
&&\qquad\qquad\qquad+\frac{4+m^2+4mx}{Y^2}\Big]f_0\,,\nn\\
f''_4&=&2\Big(\frac{x}Y+\frac{4ia}{\rho_-}\Big)f'_4+\Big[w^2a^2-\frac{\lambda+2wa(m+2x)}Y\nn\\
&&\qquad+\frac{4+m^2-4mx}{Y^2}-\frac{8iax}{Y\rho_-}+\frac{12a^2}{\rho_-^2}\Big]f_4\,,\label{eq.key.1c}\eea
where $\rho_\pm\equiv r\pm iax\,$.
Combining \eqref{eq.key.1c} with Teukolsky's master equation, one can also find that
\bea \ddot{f}_0&=&-\frac{6(r-M)}X\dot{f}_0 -\frac{U_1}{X^2}f_0\,,\nn\\
\ddot{f}_4&=&2\Big(\frac{r-M}X-\frac{4}{\rho_-}\Big)\dot{f}_4 -\Big(\frac{U_2}{X^2}-\frac{U_3}{X\rho_-^2}\Big)f_4\,,\label{eq.key.1d}\eea
where
\bea U_1&=&[w(r^2+a^2)-ma]^2+4ima(r-M)\nn\\
&&-4iMw(r^2-a^2)-(\lambda-6-4iwr)X\,,\nn\\
U_2&=&[w(r^2+a^2)-ma]^2-4ima(r-M)\nn\\
&&+4iMw(r^2-a^2)-(\lambda+4iwr)X\,,\nn\\
U_3&=&6H-4(2M+iax)\rho_--12X\,.\eea
Eqs. \eqref{eq.key.1c} and \eqref{eq.key.1d} are consistent with the fact that Teukolsky's master equation is separable.

Eq. \eqref{eq.key.1b} only contains derivatives with respect to $x$.
One can try to reduce it further by eliminating the ten mode functions, i.e., $f_{\mu\nu}\in\{f_{tt}$, $f_{tr}$, $f_{tx}$, $f_{t\phi}$, $f_{rr}$, $f_{rx}$, $f_{r\phi}$, $f_{xx}$, $f_{x\phi}$, $f_{\phi\phi}\}$, from the equations sequentially.
In principle, one can try to firstly single out a mode function, say $f_{tt}$,
then to use the \ac{DCS} procedure iteratively to lower the order of its $x$-derivatives to 0,
and then to solve $f_{tt}$ in terms of the remaining functions.
However, it has proved computationally infeasible to obtain all the mode functions from \eqref{eq.key.1b} in this way.

\subsection{Adding new constraints}

The situation can be fundamentally changed by incorporating the two additional constraints: the traceless condition \eqref{eq.traceless} and the parity constraint \eqref{eq.weyle.scalar3}.
It turns out that the combination of the equations \eqref{eq.combine24}, \eqref{eq.weyle.scalar2}, \eqref{eq.traceless} and \eqref{eq.weyle.scalar3} can be fully reduced and separated through the following steps.

\subsubsection{Applying the traceless condition}

The traceless condition \eqref{eq.traceless} is an algebraic equation of the mode functions and can be solved as
\bea f_{t\phi}&=&-\frac1{4Mar}\Big\{\Big[(r^2+a^2)^2-a^2XY\Big]f_{tt}-X^2f_{rr}\nn\\
&&\qquad\qquad-XYf_{xx}-\Big(\frac{X}Y-a^2\Big)f_{\phi\phi}\Big\}\,.\label{rst.ftphi}\eea
This eliminates one mode function from all the equations.

\subsubsection{Reducing the order of $r$-derivatives}

For the 28 partial differential equations in \eqref{eq.combine24}, \eqref{eq.weyle.scalar2} and \eqref{eq.weyle.scalar3}, one can repeat what have been done for \eqref{eq.combine24} and \eqref{eq.weyle.scalar2} in the previous subsection and reduce the order of $r$-derivatives by iterating the \ac{DCS} procedure.
In this way, the 28 equations can be reduced to
\begin{subequations}\label{eq.key.2a}
\begin{align}
&\dot{f}_{\mu\nu}=\cF^{(9)}_{\mu\nu}\Big[f_{\bullet\bullet}, f'_{\bullet\bullet}, f''_{\bullet\bullet}, f'''_{\bullet\bullet}\Big]\,, \,,\label{eq.k2a1}\\
&0=E^{(19)}\Big[f_{\bullet\bullet},f'_{\bullet\bullet},\cdots,f^{(5)}_{\bullet\bullet},f_{\bullet}, f'_{\bullet}, f''_{\bullet} \Big]\,,\label{eq.k2a2}
\end{align}
\end{subequations}
where $f_{\mu\nu},f_{\bullet\bullet}\in\{f_{tt}$, $f_{tr}$, $f_{tx}$, $f_{rr}$, $f_{rx}$, $f_{r\phi}$, $f_{xx}$, $f_{x\phi}$, $f_{\phi\phi}\}$, and $f_{\bullet} \in\{f_0, f_4,\td{f}_0, \td{f}_4\}$.
There are 9 equations in \eqref{eq.k2a1} and 19 in \eqref{eq.k2a2}.

\subsubsection{Reducing the order of $x$-derivatives}

By giving numerical values to the physical parameters, one can further reduce \eqref{eq.k2a2} to 13 equations, among which nine are of the form:
\begin{subequations}\label{eq.key.2b}
\begin{align}
f'_{\mu\nu}&=\cF^{(6)}_{\mu\nu}\Big[f_{\bullet\bullet},f'_{\bullet\bullet},f_{\bullet}, f'_{\bullet}\Big]\,,\label{eq.k2b1}\\
f''_{\mu\nu}&=\cF^{(3)}_{\mu\nu}\Big[f_{\bullet\bullet},f'_{\bullet\bullet},f_{\bullet}, f'_{\bullet}\Big]\,,\label{eq.k2b2}
\end{align}
\end{subequations}
where for \eqref{eq.k2b1}, $f_{\mu\nu}\in\{f_{tt}$, $f_{tx}$, $f_{rx}$, $f_{r\phi}$, $f_{x\phi}$, $f_{\phi\phi}\}$,
and for \eqref{eq.k2b2}, $f_{\mu\nu}\in\{f_{tr}$, $f_{rr}$, $f_{xx}\}$.
So, there are six (three) equations having $x$-derivatives up to the first (second) order.
Here, it is not of fundamental importance which functions are assigned the second-order derivatives and which the first-order ones, but the current choice appears convenient.
Note the number of equations in  \eqref{eq.key.2b} equals the number of remaining mode functions.

The remaining 4 of the 13 equations are \eqref{eq.key.1c} and their parity conjugates, which can be obtained by acting on \eqref{eq.key.1c} with the parity operator.
By acting on \eqref{eq.key.1d} with the parity operator, one can also obtain the corresponding equations for $\td{f}_0$ and $\td{f}_4$.

\subsubsection{Solving for the mode functions}

There are significant improvements of \eqref{eq.key.2b} over \eqref{eq.key.1b}.
Firstly, only three mode functions in \eqref{eq.key.2b} have second-order derivatives, while the remaining 6 only have first-order derivatives.
Secondly, the overall expressions of \eqref{eq.key.2b} are much simpler than those of \eqref{eq.key.1b}.
Consequently, all mode functions can now be solved sequentially from \eqref{eq.key.2b} by iterating the \ac{DCS} procedure.

The process can go largely unobstructed until there are only two equations left, which can be schematically written as
\bea E^{(2)}\Big[f_{\ast\ast},f'_{\ast\ast},\cdots,f^{(6)}_{\ast\ast}, f_{\bullet}, f'_{\bullet}\Big]&=&0\,,\label{eq.key.3a}\eea
which contain two unsolved mode functions, $f_{\ast\ast}\in\{f_{tt}$, $f_{xx}\}$.
(Again, it is not a unique but convenient choice to keep $f_{tt}$ and $f_{xx}$ as last two unsolved mode functions.)
Simply because the expressions have become too lengthy, the difficulty increases substantially when one attempts to solve for yet another mode function, say $f_{tt}$, from the equations.
To overcome the difficulty, one can let the coordinate $r$ take a numerical value as well.
This is possible because $r$ is not needed for the derivatives.\footnote{In fact, one can do this right at the beginning of reducing \eqref{eq.key.2b}, but will pay the price of having more difficulty to recover the $r$-dependence in the end.}
In this way, $f_{tt}$ can be solved and one is left with a single equation containing one unknown function, i.e.,
\bea E^{(1)}\Big[f_{xx},f'_{xx},\cdots,f^{(12)}_{xx}, f_{\bullet}, f'_{\bullet}\Big]=0\,.\label{eq.key.3b}\eea

Because there is only one equation left, the \ac{DCS} procedure can no longer be applied.
In addition, since \eqref{eq.key.3b} contains derivatives of $f_{xx}$ up to the 12th-order and is very complicated even when all the physical parameters and $r$ have taken numerical values, there is no hope to solve the equation through integration.
So, the only way to solve the equation appears to be via an ansatz.

The ansatz can be constructed as follows.
Since \eqref{eq.key.3b} only depends on $f_{xx}$, $f_i\in\{f_0,f_4,\td{f}_0,\td{f}_4\}$, and their derivatives, it can be expected that $f_{xx}$ must be a combination of the $f_i$'s and their derivatives.
What's more, the corresponding coefficients must be finite order polynomials in $x$ both in the numerator and denominator,
\bea f_{xx}=\frac1{A(x)}\sum_i\Big[p_i(x)f_i+q_i(x)f'_i\Big]\,,\label{ansatz.fxx0}\eea
where $A(x)$, $p_i(x)$ and $q_i(x)$ are finite order polynomials in $x$.
Given an assumed degree for each of the polynomials, the task is then to determine all the unknown coefficients in the polynomials by plugging \eqref{ansatz.fxx0} into \eqref{eq.key.3b}.
However, this typically leads to nonlinear equations for the unknown coefficients and are impossible to solve.
The only case when one can get linear equations for the unknown coefficients is when the polynomial in the denominator, i.e. $A(x)$, is known.
Given the Kerr metric \eqref{metric.kerr}, a natural choice for $A(x)$ is some product of $Y$ and $H$.
For this reason, one can try the following ansatz,
\bea f_{xx}=\frac1{(HY)^n}\sum_i\Big[p_i(x)f_i+q_i(x)f'_i\Big]\,,\label{ansatz.fxx1}\eea
where $n$ takes a positive integer value, and the denominator depends on $x$ only through $H$ and $Y$.

The $p_i$'s and $q_i$'s can be solved by plugging \eqref{ansatz.fxx1} into \eqref{eq.key.3b}, using \eqref{eq.key.1c} and its parity conjugate to lower the derivatives of $f_i$'s to the first order, and then assuming that all the resultant coefficients of the $f_i$'s and $f'_i$'s are zero.
By experimenting with $n$ as large as 6, a unique solution can be found.
After cancelling with factors in the resultant $p_i$'s and $q_i$'s, it turns out that none of the terms in $f_{xx}$ contains power of $H$ (or $Y$) higher than 1 (or 3) in the denominator.

Given $f_{xx}$, one can solve for other mode functions through the following steps:
\begin{itemize}
\item Firstly, one can obtain $f_{tt}$ directly as it has been solved in terms of $f_{xx}$ when reducing \eqref{eq.key.3a} to \eqref{eq.key.3b}.
\item Secondly, treating all the coefficients in the $x$ polynomials in $f_{tt}$ and $f_{xx}$ as unknown constants and plugging the expressions into \eqref{eq.key.3a}, one can restore the $r$-dependence of both $f_{tt}$ and $f_{xx}$.
\item Thirdly, since all the other functions, $f_{tr}$, $f_{tx}$, $f_{rr}$, $f_{rx}$, $f_{r\phi}$, $f_{x\phi}$ and $f_{\phi\phi}$, have been solved in terms of $f_{tt}$ and $f_{xx}$ during the reduction of \eqref{eq.key.2b} to \eqref{eq.key.3a}, their expressions can also be obtained without trouble.
\end{itemize}
Together with \eqref{rst.ftphi}, one obtains explicit expressions relating all the ten mode functions from the perturbed metric, i.e. $f_{\mu\nu} \in\{f_{tt}$, $f_{tr}$, $f_{tx}$, $f_{t\phi}$, $f_{rr}$, $f_{rx}$, $f_{r\phi}$, $f_{xx}$, $f_{x\phi}$, $f_{\phi\phi}\}$, to the four mode functions from the perturbed Weyl scalars, i.e., $f_i\in\{f_0, f_4,\td{f}_0, \td{f}_4\}$.

The next step is to restore the dependence on the physical parameters.
Since the functional structure of all the mode functions are now clear, one can make an ansatz of the mode functions by replacing all the numbers in their expressions with a set of unknown coefficients, and then plug into \eqref{eq.k2a2}.
The unknown coefficients can then be found and the dependence of the mode functions on the physical parameters can be restored.

The final step is to check the results against \eqref{eq.k2a1}, which has not been used so far.
Remarkably, it turns out that the only extra constraint imposed by \eqref{eq.k2a1} is a fixed interdependence among $f_0$, $f_4$, $\td{f}_0$ and $\td{f}_4$.

\section{Results}\label{sec:rst}

In this section, the key results of the decouple mode functions, $f_{\mu\nu}(r,x)$, will be presented.

In \cite{Mei:2023pho}, it has been noticed that the solutions found are rather lengthy.
At that time, it was suspected that the lack of a direct derivation was the reason.
Here, with the direct derivation available, the results still turn out to be not so simple.
In fact, as will be shown below, one actually recovers the solutions in \cite{Mei:2023pho} as a byproduct of the derivation.
Thus, it may be unavoidable that the analytical expressions for the perturbations of a Kerr black hole are not as simple as one might hope.
Of course, one can never rule out the possibility that there might be a better way to separate the equations which can produce much simpler results.

In any case, for the purpose of this work, several sets of finite order polynomials in $r$ and $x$ will be introduced to help present the results, including
$A_i$, $i\in\{1,\cdots,40\}$,
$B_i$, $i\in\{1,2,3,4\}$,
$C_i$, $i\in\{1,2,\cdots,6\}$,
$D_i$ and $E_i$, $i\in\{1,\cdots,40\}$, in
\eqref{rst.metric1b}, \eqref{rst.p0p4},\eqref{rst.p0p4-2}, \eqref{rst.metric2b} and \eqref{rst.metric2c}, respectively.
All of them will be given in separate supplemental files.

By following the steps explained in the previous section, but without using \eqref{eq.k2a1}, the mode functions can be found as:
\bea f_a=\check{f}_{a}+\epsilon\cP[\check{f}_{a}]\,,\quad f_b=\check{f}_{b}-\epsilon\cP[\check{f}_{b}]\,,\label{rst.metric1a}\eea
where $a\in\{{tt}$, ${tr}$, ${t\phi}$, ${rr}$, ${r\phi}$, ${xx}$, ${\phi\phi}\}$, $b\in\{{tx}$, ${rx}$, ${x\phi}\}$, and
\begin{widetext}
\bea \check{f}_{tt} &=&\frac{A_{20} X^2 f_0 -4i A_{30} \rho_-^3 f_4 - 2 A_4 X^2 Y  f'_0 + 8 A_3 \rho_-^4 Y  f'_4}{24 w^2 f_r \rho_-^3 \rho_+^3 Y}\,,\nn\\
\check{f}_{tr} &=&\frac{ A_{21} X^2 f_0 +4i A_{33} \rho_-^3 f_4 - 2 A_{5} X^2 Y  f'_0 - 8 A_{6} \rho_-^4 Y  f'_4}{24 w^2 f_r \rho_-^2 \rho_+^2 X     Y}\,,\nn\\
\check{f}_{tx} &=&\frac{ i A_{19} X^2 f_0 -4i A_{26} \rho_-^3 f_4 - A_{1} X^2 Y  f'_0 + 4 A_{2} \rho_-^4 Y  f'_4}{12 w^2 f_r \rho_-^2 \rho_+^2     Y^2}\,,\nn\\
\check{f}_{t\phi} &=&-\frac{A_{31} X^2 f_0 -4i A_{36} \rho_-^3 f_4 - 2 A_{15} X^2 Y  f'_0 + 8 A_{16} \rho_-^4 Y  f'_4}{24 w^2 f_r \rho_-^3 \rho_+^3 Y}\,,\nn\\
\check{f}_{rr} &=& \frac{ A_{29} X^2 f_0 -4i A_{38} \rho_-^3 f_4 - 2 A_{13} X^2 Y  f'_0 + 8 A_{14} \rho_-^4 Y  f'_4}{24 w^2 f_r \rho_- \rho_+ X^2     Y}\,,\nn\\
\check{f}_{rx} &=& \frac{ i A_{25} X^2 f_0 +4 A_{35} \rho_-^3 f_4 - i A_{12} X^2 Y  f'_0 + 4i A_{11} \rho_-^4 Y  f'_4}{24 w^2 f_r \rho_- \rho_+ X     Y^2}\,,\nn\\
\check{f}_{r\phi} &=& -\frac{A_{28} X^2 f_0 +4i A_{37} \rho_-^3 f_4 - 2 A_{17} X^2 Y  f'_0 - 8 A_{18} \rho_-^4 Y  f'_4}{24 w^2 f_r \rho_-^2 \rho_+^2 X     Y}\,,\nn\\
\check{f}_{xx} &=& -\frac{A_{22} X^2 f_0 -4i A_{32} \rho_-^3 f_4 - 2 A_{8} X^2 Y  f'_0 + 8 A_{7} \rho_-^4 Y  f'_4}{24 w^2 f_r \rho_- \rho_+ Y^3}\,,\nn\\
\check{f}_{x\phi} &=& \frac{ A_{27} X^2 f_0 -4i A_{34} \rho_-^3 f_4 - A_{10} X^2 Y  f'_0 + 4 A_{9} \rho_-^4 Y  f'_4}{12 w^2 f_r \rho_-^2 \rho_+^2     Y^2}\,,\nn\\
\check{f}_{\phi\phi} &=& \frac{ A_{39} X^2 f_0 -4i A_{40} \rho_-^3 f_4 - 2 A_{23} X^2 Y  f'_0 + 8 A_{24} \rho_-^4 Y  f'_4}{24 w^2 f_r \rho_-^3 \rho_+^3 Y}\,.
\label{rst.metric1b}\eea
Here
\bea f_r=2\Big[(r^2 + a^2) w - a m\Big]^3 + 2 w (a^2 - M r) X + \Big[(r^2 + a^2) w - a m\Big]\Big[2 (r - M)^2 -  \lambda X\Big] \,.\eea

Plugging the above results into \eqref{eq.k2a1}, one can find that the only extra constraint is a fixed interdependence among $f_0$, $f_4$, $\td{f}_0$ and $\td{f}_4$,
\bea \dot{f}_0&=&\frac{B_2 f_0}{4 f_r X}
+\frac{B_3 \rho_-^3 f_4}{f_r X Y^2}
+\frac{12 M w \rho_+^4\td{f}_4}{f_r X}
+\frac{4 i f_x \rho_-^4 f'_4}{f_r X Y}\,,\nn\\
\dot{f}_4 &=&\frac{-i B_1 X^3 f_0}{16 f_r \rho_-^4 Y^2}
+\frac{B_4 f_4}{4 f_r \rho_- X}
+\frac{3 M w X^3 \td{f}_0}{4 f_r \rho_-^4}
+\frac{i f_x X^3 f'_0}{4f_r \rho_-^4 Y}\,,
\label{rst.p0p4}\eea
where the equations of $\dot{\td{f}}_0$, and $\dot{\td{f}}_4$ can be obtained by acting on \eqref{rst.p0p4} with the parity operator, and
\bea f_x=2m(m^2-x^2)-(\lambda+6amw)(m-awY)Y-2aw(1+a^2w^2Y)Y^2\,.\eea
Using these equations, one can solve $f_0$ in terms of $f_4$ and $\td{f}_4$, and vice versa,
\begin{subequations}\label{rst.p0p4-2}
\begin{align}
f_0 &= \frac{-4 \rho_-^4}{Q X^3}
\Big[\frac{C_6 f_4}{\rho_-^2 X Y^2}
-\frac{4 C_3 f_r \dot{f}_4}{\rho_- Y^2}
+\frac{4 i C_5 f_x f'_4}{\rho_- X Y}
-\frac{16 i f_r f_x \dot{f}'_4}{Y}
+\frac{12 M w C_4 \rho_+^3 \td{f}_4}{\rho_-^4 X}
-\frac{48 M w f_r \rho_+^4 \dot{\td{f}}_4}{\rho_-^4}\Big]\,,\label{rst.p0p4-2a}\\
f_4 &= \frac{i f_r f_x X}{4Q \rho_-^4}
\Big[\frac{C_1 C_2 f_0}{f_r f_x X Y^2}
-\frac{4 C_1 \dot{f}_0}{f_x Y^2}
-\frac{4 C_2 f'_0}{f_r X Y}
+\frac{16 \dot{f}'_0}{Y}
+\frac{12 i M w C_2 \td{f}_0}{f_r f_x X}
-\frac{48 i M w \dot{\td{f}}_0}{f_x}\Big]\,,\label{rst.p0p4-2b}
\end{align}
\end{subequations}
where
\bea Q=\lambda^2(\lambda-2)^2+8wa(\lambda-2)(5\lambda-4)(m-aw)+48(wa)^2[2\lambda-4+3(m-wa)^2]+(12Mw)^2\,.\eea

With \eqref{rst.p0p4-2}, one can express the mode functions either purely in terms of $f_0$ and $\td{f}_0$, or equivalently, purely in terms of $f_4$ and $\td{f}_4$.
For example, if one plugs \eqref{rst.p0p4-2a} into \eqref{rst.metric1b}, then the mode functions can be expressed purely in terms of $f_4$ and $\td{f}_4$.
Remarkably, the resultant mode functions can be written in the following form:
\bea f_a&=&f^{(4)}_a+f^{(0)}_a+\epsilon\cP[f^{(4)}_a+f^{(0)}_a]\,,\nn\\
f_b&=&f^{(4)}_b+f^{(0)}_b-\epsilon\cP[f^{(4)}_b+f^{(0)}_b]\,,\label{rst.metric2a}\eea
where $a\in\{{tt}$, ${tr}$, ${t\phi}$, ${rr}$, ${r\phi}$, ${xx}$, ${\phi\phi}\}$, and
$b\in\{{tx}$, ${rx}$, ${x\phi}\}$.
The components $f^{(4)}_{\mu\nu}$ and $f^{(0)}_{\mu\nu}$ are given by:
\bea f^{(4)}_{tt}&=&\frac{-i D_{32} f_4}{18M w^3 \rho_- \rho_+^3 X^2 Y}+\frac{D_{13} f'_4}{9 M w^3 \rho_+^3 X^2}-\frac{iD_{12} \dot{f}_4}{18M w^3 \rho_+^3 X Y}+\frac{D_{2} \rho_- \dot{f}'_4}{9 M w^3 \rho_+^3 X}\,,\nn\\
f^{(4)}_{tr}&=&\frac{-i D_{33} f_4}{18M w^3 \rho_+^2 X^3 Y}+\frac{D_{17} \rho_- f'_4}{9 M w^3 \rho_+^2 X^3}-\frac{i D_{16} \rho_- \dot{f}_4}{18M w^3 \rho_+^2 X^2 Y}+\frac{D_{3} \rho_-^2 \dot{f}'_4}{9 M w^3 \rho_+^2 X^2}\,,\nn\\
f^{(4)}_{tx}&=&\frac{i D_{31} f_4}{9M w^3 \rho_+^2 X^2 Y^2}-\frac{D_{15} \rho_- f'_4}{18 M w^3 \rho_+^2 X^2 Y}+\frac{i D_{11} \rho_- \dot{f}_4}{9M w^3 \rho_+^2 X Y^2}-\frac{D_{1} \rho_-^2 \dot{f}'_4}{9 M w^3 \rho_+^2 X Y}\,,\nn\\
f^{(4)}_{t\phi}&=&\frac{i D_{38} f_4}{18M w^3 \rho_- \rho_+^3 X^2 Y}-\frac{D_{27} f'_4}{9 M w^3 \rho_+^3 X^2}+\frac{i D_{22} \dot{f}_4}{18M w^3 \rho_+^3 X Y}-\frac{D_{8} \rho_- \dot{f}'_4}{9 M w^3 \rho_+^3 X}\,,\nn\\
f^{(4)}_{rr}&=&\frac{-i D_{36} f_4 \rho_-}{18M w^3 \rho_+ X^4 Y}+\frac{D_{21} \rho_-^2 f'_4}{18 M w^3 \rho_+ X^4}-\frac{i D_{25} \rho_-^2 \dot{f}_4}{18M w^3 \rho_+ X^3 Y}+\frac{D_{7} \rho_-^3 \dot{f}'_4}{9 M w^3 \rho_+ X^3}\,,\nn\\
f^{(4)}_{rx}&=&-\frac{D_{35} f_4 \rho_-}{9 M w^3 \rho_+ X^3 Y^2}-\frac{i D_{23} \rho_-^2 f'_4}{18M w^3 \rho_+ X^3 Y}-\frac{D_{20} \rho_-^2 \dot{f}_4}{18 M w^3 \rho_+ X^2 Y^2}-\frac{i D_{6} \rho_-^3 \dot{f}'_4}{18M w^3 \rho_+ X^2 Y}\,,\nn\\
f^{(4)}_{r\phi}&=&\frac{i D_{39} f_4}{18M w^3 \rho_+^2 X^3 Y}-\frac{D_{28} \rho_- f'_4}{9 M w^3 \rho_+^2 X^3}+\frac{i D_{26} \rho_- \dot{f}_4}{18M w^3 \rho_+^2 X^2 Y}-\frac{D_{9} \rho_-^2 \dot{f}'_4}{9 M w^3 \rho_+^2 X^2}\,,\nn\\
f^{(4)}_{xx}&=&\frac{i D_{34} f_4 \rho_-}{18M w^3 \rho_+ X^2 Y^3}-\frac{D_{19} \rho_-^2 f'_4}{18 M w^3 \rho_+ X^2 Y^2}+\frac{i D_{14} \rho_-^2 \dot{f}_4}{18M w^3 \rho_+ X Y^3}-\frac{D_{4} \rho_-^3 \dot{f}'_4}{9 M w^3 \rho_+ X Y^2}\,,\nn\\
f^{(4)}_{x\phi}&=&\frac{i D_{37} f_4}{18M w^3 \rho_+^2 X^2 Y^2}-\frac{D_{24} \rho_- f'_4}{18 M w^3 \rho_+^2 X^2 Y}+\frac{i D_{18} \rho_- \dot{f}_4}{9M w^3 \rho_+^2 X Y^2}-\frac{D_{5} \rho_-^2 \dot{f}'_4}{9 M w^3 \rho_+^2 X Y}\,,\nn\\
f^{(4)}_{\phi\phi}&=&\frac{-i D_{40} f_4}{18M w^3 \rho_- \rho_+^3 X^2 Y}+\frac{D_{30} f'_4}{18 M w^3 \rho_+^3 X^2}-\frac{i D_{29} \dot{f}_4}{18M w^3 \rho_+^3 X Y}+\frac{D_{10} \rho_- \dot{f}'_4}{9 M w^3 \rho_+^3 X}\,,\label{rst.metric2b}\eea
and, with $f_0$ determined by \eqref{rst.p0p4-2a} but without the $\td{f}_4$ terms,
\bea f^{(0)}_{tt}&=&\frac{E_{23} f_0}{72 M w^3 \rho_-^3 \rho_+^3 Y}-\frac{E_{10} f'_0}{72 M w^3 \rho_-^3 \rho_+^3 Y}+\frac{E_{15} \dot{f}_0}{72 M w^3 \rho_-^3 \rho_+^3 Y}-\frac{E_{2} \dot{f}'_0}{72 M w^3 \rho_-^3 \rho_+^3 Y}\,,\nn\\
f^{(0)}_{tr}&=&-\frac{E_{30} f_0}{72 M w^3 \rho_-^2 \rho_+^2 X Y}+\frac{E_{13} f'_0}{72 M w^3 \rho_-^2 \rho_+^2 X Y}-\frac{E_{16} \dot{f}_0}{72 M w^3 \rho_-^2 \rho_+^2 X Y}+\frac{E_{3} \dot{f}'_0}{72 M w^3 \rho_-^2 \rho_+^2 X Y}\,,\nn\\
f^{(0)}_{tx}&=&\frac{-i E_{21} f_0}{72M w^3 \rho_-^2 \rho_+^2 Y^2}+\frac{E_{6} f'_0}{72 M w^3 \rho_-^2 \rho_+^2 Y^2}+\frac{i E_{12} \dot{f}_0}{72M w^3 \rho_-^2 \rho_+^2 Y^2}+\frac{E_{1} \dot{f}'_0}{72 M w^3 \rho_-^2 \rho_+^2 Y^2}\,,\nn\\
f^{(0)}_{t\phi}&=&\frac{-E_{36} f_0}{72 M w^3 \rho_-^3 \rho_+^3 Y}+\frac{E_{22} f'_0}{72 M w^3 \rho_-^3 \rho_+^3 Y}-\frac{E_{32} \dot{f}_0}{72 M w^3 \rho_-^3 \rho_+^3 Y}+\frac{E_{9} \dot{f}'_0}{72 M w^3 \rho_-^3 \rho_+^3 Y}\,,\nn\\
f^{(0)}_{rr}&=&\frac{E_{34} f_0}{72 M w^3 \rho_- \rho_+ X^2 Y}-\frac{E_{20} f'_0}{72 M w^3 \rho_- \rho_+ X^2 Y}+\frac{E_{25} \dot{f}_0}{72 M w^3 \rho_- \rho_+ X^2 Y}-\frac{E_{8} \dot{f}'_0}{72 M w^3 \rho_- \rho_+ X^2 Y}\,,\nn\\
f^{(0)}_{rx}&=&\frac{-i E_{33} f_0}{72M w^3 \rho_- \rho_+ X Y^2}+\frac{i E_{19} f'_0}{72M w^3 \rho_- \rho_+ X Y^2}-\frac{i E_{24} \dot{f}_0}{72M w^3 \rho_- \rho_+ X Y^2}+\frac{i E_{7} \dot{f}'_0}{72M w^3 \rho_- \rho_+ X Y^2}\,,\nn\\
f^{(0)}_{r\phi}&=&\frac{E_{38} f_0}{72 M w^3 \rho_-^2 \rho_+^2 X Y}-\frac{E_{27} f'_0}{72 M w^3 \rho_-^2 \rho_+^2 X Y}+\frac{E_{29} \dot{f}_0}{72 M w^3 \rho_-^2 \rho_+^2 X Y}-\frac{E_{11} \dot{f}'_0}{72 M w^3 \rho_-^2 \rho_+^2 X Y}\,,\nn\\
f^{(0)}_{xx}&=&\frac{-E_{26} f_0}{72 M w^3 \rho_- \rho_+ Y^3}+\frac{E_{14} f'_0}{72 M w^3 \rho_- \rho_+ Y^3}-\frac{E_{18} \dot{f}_0}{72 M w^3 \rho_- \rho_+ Y^3}+\frac{E_{4} \dot{f}'_0}{72 M w^3 \rho_- \rho_+ Y^3}\,,\nn\\
f^{(0)}_{x\phi}&=&\frac{-E_{35} f_0}{72 M w^3 \rho_-^2 \rho_+^2 Y^2}+\frac{E_{17} f'_0}{72 M w^3 \rho_-^2 \rho_+^2 Y^2}-\frac{E_{31} \dot{f}_0}{72 M w^3 \rho_-^2 \rho_+^2 Y^2}+\frac{E_{5} \dot{f}'_0}{72 M w^3 \rho_-^2 \rho_+^2 Y^2}\,,\nn\\
f^{(0)}_{\phi\phi}&=&\frac{E_{40} f_0}{72 M w^3 \rho_-^3 \rho_+^3 Y}-\frac{E_{37} f'_0}{72 M w^3 \rho_-^3 \rho_+^3 Y}+\frac{E_{39} \dot{f}_0}{72 M w^3 \rho_-^3 \rho_+^3 Y}-\frac{E_{28} \dot{f}'_0}{72 M w^3 \rho_-^3 \rho_+^3 Y}\,.\label{rst.metric2c}\eea
\end{widetext}

The above calculation can be repeated for \eqref{rst.p0p4-2b}, and as expected, one can get exactly the same results as in \eqref{rst.metric2a}, \eqref{rst.metric2b} and \eqref{rst.metric2c}.
More explicitly, by plugging the whole expression of \eqref{rst.p0p4-2b} into \eqref{rst.metric1b}, one can express the mode functions purely in terms of $f_0$ and $\td{f}_0$.
The result can be cast into the form \eqref{rst.metric2a}, for which the $f_4$ in \eqref{rst.metric2b} is given by \eqref{rst.p0p4-2b} but without the $\td{f}_0$ terms.

It is remarkable that, if treating $f_0$ and $f_4$ as mutually independent functions, $f^{(4)}_{\mu\nu}$ and $f^{(0)}_{\mu\nu}$ are actually independent solutions of \eqref{eq.huv}, \eqref{eq.gauge.DD}, \eqref{eq.eigen.guv}, \eqref{eq.weyle.scalar2} and \eqref{eq.traceless}.
One can further check that they are also identical to the two solutions found in \cite{Mei:2023pho}.
In \cite{Mei:2023pho}, one has to make an ansatz for all the mode functions to obtain the solutions.
Here, they naturally appear as a byproduct of the derivation.

\section{Summary and discussion}\label{sec:sum}

In this work, an explicit derivation has been presented to separate the \acp{LEE} in Kerr.
The derivation has been made possible by incorporating the following equations and constraints:
\begin{itemize}
\item the \acp{LEE} in Kerr, \eqref{eq.huv};
\item the de Donder gauge condition, \eqref{eq.gauge.DD};
\item the eigen-equation of the \ac{KY} operator, \eqref{eq.eigen.guv};
\item the fixed parity assumption and the Weyl scalar constraints, \eqref{def.parity.fuv}, \eqref{eq.weyle.scalar2} and \eqref{eq.weyle.scalar3}; and
\item the traceless condition, \eqref{eq.traceless}.
\end{itemize}
All these equations and constraints are physically well motivated and are obviously mutually compatible.

With these equations, explicit formulas have been derived, for the first time, expressing the linear order metric perturbation entirely in terms of some master functions, i.e., $f_4$ and $\td{f}_4$ (or, equivalently, $f_0$ and $\td{f}_0$).
These master functions satisfy the readily separable equations, \eqref{eq.key.1c}, \eqref{eq.key.1d} and their parity conjugates, all of which are variants of Teukolsky's master equation.

The derivation also leads to a fixed relation between $f_0$ and $f_4$, i.e. \eqref{rst.p0p4-2}, that resembles the Teukolsky-Starobinsky identities.
A notable feature of \eqref{rst.p0p4-2} is that the results are valid only when $Q\neq0$.
In this case, $\psi_0=0$ means $\psi_4=0$, and vice versa.
This is consistent with what is expected for physically well behaved perturbations \cite{Wald:1973wwa}.
As one may expect, $Q$ is nothing but the Teukolsky-Starobinsky constant, which is the same as that in, e.g.,
(2.23) of \cite{Berens:2024czo}, with $\lambda_{w\ell m}^{(s)}=\lambda-s(s+1)$, and
(7) of \cite{Chandrasekhar:1984mgh}, with $\lambdabar=\lambda-2$ and $\sigma^+=-w$.

In the case when $Q=0$, one can get the so called algebraically special solutions for which only one of $\psi_0$ and $\psi_4$ is non-vanishing \cite{Chandrasekhar:1984mgh}.
This appears to indicate that $Q=0$ is a necessary condition for the existence of algebraically special solutions.
However, in the context of the present calculation, if one drops the requirement of fixed parities for the mode functions, and hence also \eqref{eq.weyle.scalar3}, then there can be no fixed relation between $f_0$ and $f_4$.
In this case, algebraically special solutions can exist even when $Q\neq0$.
As concrete examples, one can check that $f^{(4)}_{\mu\nu}$ in \eqref{rst.metric2b} and $f^{(0)}_{\mu\nu}$ \eqref{rst.metric2c} have the following properties:
\bea \cF_0[f^{(4)}_{\mu\nu}]=0\,,\quad \cF_4[f^{(4)}_{\mu\nu}]=f_4\,,\nn\\
\cF_0[f^{(0)}_{\mu\nu}]=f_0\,,\quad \cF_4[f^{(0)}_{\mu\nu}]=0\,,\eea
where $f_0$ and $f_4$ are treated as mutually independent functions, and $\cF_0[\cdots]$ and $\cF_4[\cdots]$ are defined in \eqref{eq.weyle.scalar2}.

As an outlook, the present work can be extend in several directions.
Firstly, since \eqref{eq.key.3b} has been solved by using an ansatz, i.e. \eqref{ansatz.fxx1}, some generality might have been lost.
It will be interesting to see if there is a more general solution than the one found here.
Secondly, although there is encouraging sign suggesting that the mode functions found can describe physically well behaved perturbations, i.e. by having a fixed relation between $f_0$ and $f_4$ \cite{Wald:1973wwa}, the properties and true physical relevance of the mode functions require more extensive study to clarify.
Thirdly, this work has only considered the vacuum case, but most physically interesting problems involve sources.
So, it will be of great importance to extend the present derivation to the non-vacuum cases.

\section*{Acknowledgments}

The work has been supported in part by the National Key Research and Development Program of China (Grant No. 2023YFC2206700), and the Fundamental Research Funds for the Central Universities, Sun Yat-sen University.

\bibliographystyle{apsrev4-1}
\bibliography{ref}

\end{document}